\documentstyle[epsfig]{qcdparis}

   \def\unlock{\catcode`@=11}

   \unlock

   \def\gsim{\mathrel{\mathpalette\@versim>}}
   \def\lsim{\mathrel{\mathpalette\@versim<}}

   \def\@versim#1#2{\vcenter{\offinterlineskip
        \ialign{$\m@th#1\hfil##\hfil$\crcr#2\crcr\sim\crcr } }}

\newcommand{\beq}{\begin{equation}}
\newcommand{\eeq}{\end{equation}}
\newcommand{\bea}{\begin{eqnarray}}
\newcommand{\eea}{\end{eqnarray}}
\newcommand{\kt}{k_\perp}
\newcommand{\kta}{k_{a\perp}}
\newcommand{\ktb}{k_{b\perp}}
\newcommand{\kti}{k_{i\perp}}
\newcommand{\qt}{q_\perp}
\newcommand{\ra}{\rightarrow}

\renewcommand{\a}{\alpha}
\newcommand{\s}{\sigma}
\newcommand{\hsig}{\hat \sigma}
\newcommand{\hs}{\hat s}
\newcommand{\th}{\hat t}
\newcommand{\nn}{\nonumber}

\begin{document}
\pagestyle{plain}
\title{Forward-jet production in DIS}

\author{Vittorio Del Duca}

\affil{Deutsches Elektronen-Synchrotron \\
DESY, D-22603 Hamburg , GERMANY}

\abstract{In this talk I report on some work done in collaboration
with Jochen Bartels and Mark W\"usthoff on forward-jet inclusive production
in DIS in the small-$x_{bj}$ limit. We work in the HERA lab frame and
examine the jet production rate and the lepton-jet azimuthal-angle
correlation at small values of $x_{bj}$.}

\resume{ }

\twocolumn[\maketitle]
\fnm{7}{Talk given at the Workshop on Deep Inelastic Scattering and QCD,
Paris, April 1995}

The BFKL theory \cite{lip}
describes the dynamics of a short-distance strong-interaction
process in the limit of high squared parton c.m. energy $\hs$ and fixed
momentum transfer $\th$, by computing the scattering amplitude with
exchange of a color-singlet two-gluon ladder in the crossed channel.
At $\th=0$ the amplitude is related via the $\hs$-channel unitarity
to the total cross section with exchange of a one-gluon ladder. The
final-state gluons obtained by cutting the ladder
obey the multi-Regge kinematics, i.e. they are strongly ordered in the
rapidity $\eta$ and have comparable transverse momentum, of size $\kt$,
\beq
\eta_1 \gg ...\gg \eta_m;\qquad \kti\simeq\kt\, .\label{mreg}
\eeq
The exchange of a one-gluon cut ladder (\fref{uno}) is described
by the function,
\ffig{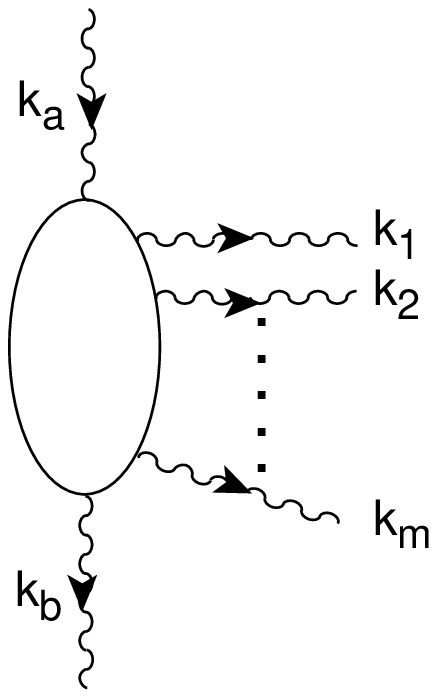}{40mm}{\em The one-gluon cut ladder.}{uno}
\bea
& & f(\kta,\ktb,\tilde{\phi},\eta)\, =\, {1\over (2\pi)^2}\, {1\over (\kta^2
\ktb^2)^{1/2}}\label{solc}\\ &\times& \sum_{n=-\infty}^{\infty}
e^{in\tilde{\phi}}\, \int_{-\infty}^{\infty} d\nu\,
e^{\omega(\nu,n)\eta}\, e^{i\nu\ln(\kta^2/\ktb^2)}\, ,\nn
\eea
with $\bf\kta$ and $\bf\ktb$ the transverse momenta of the gluons
coming in and going out of the ladder, $\tilde\phi$ the azimuthal angle
between them, $\eta\simeq\ln(\hs/\kt^2)$ an evolution parameter of
the ladder required to be large, and
\beq
\omega(\nu,n)\, =\, -2{\a_s N_c\over\pi}\, {\rm Re}\left[\psi\left({|n|+1
\over 2} +i\nu\right) -\psi(1)\right]\, ,\label{om}
\eeq
with $\psi$ the logarithmic derivative of the $\Gamma$ function.

In fully inclusive
DIS the parameter $\eta$ is related to the Biorken variable $x_{bj}$,
$\eta=\ln(1/x_{bj})$, thus evidence of the BFKL dynamics is searched
in the small-$x_{bj}$ evolution of the $F_2$ structure function \cite{cat};
in inclusive two-jet production in hadron-hadron collisions
$\eta$ is the rapidity difference between the tagging jets,
$\eta=\eta_{j_1}-\eta_{j_2}$, \cite{MN,DS}, and accordingly
evidence of the BFKL dynamics is searched in two-jet events at large
rapidity intervals \cite{d0}.
A process which encompasses features of both the processes outlined above
is forward-jet production in DIS (\fref{due}) \cite{muel,bdl}. In this case
$\eta$ is related to $x_{bj}$ and to the momentum fraction $x$ of the
parton initiating the hard scattering through $\eta=\ln(x/x_{bj})$.
\ffig{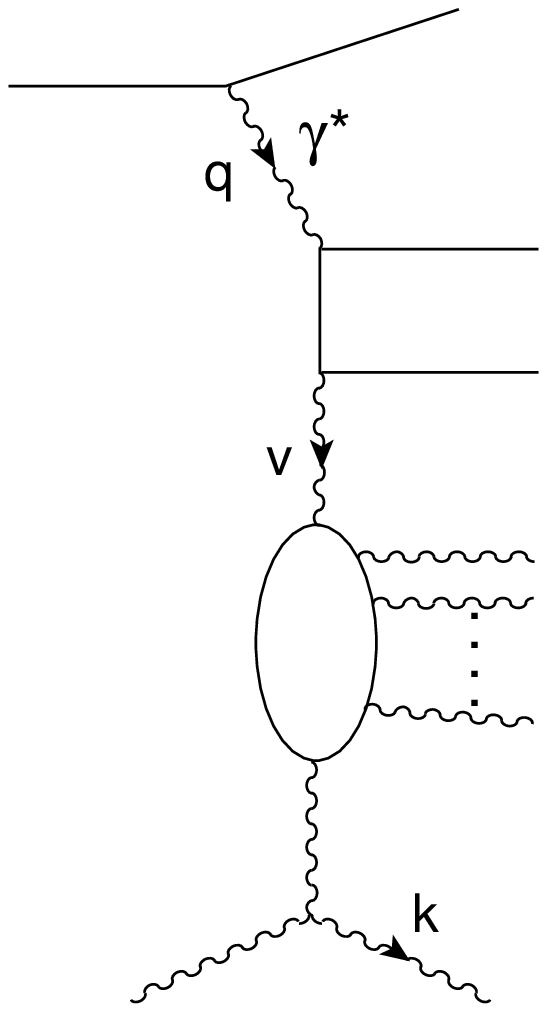}{60mm}{\em Forward-jet production in DIS.}{due}
Producing the jet forward, i.e. with a small scattering angle with
respect to the proton beam, ensures that $x$ is not small; $\eta$ is then
made large by selecting events at small $x_{bj}$.

The function $f$ (\ref{solc}) induces a strong enhancement in the total
parton cross section when $\eta$ grows \cite{lip}. In a hadron collider
$\eta=\eta_{j_1}-\eta_{j_2}\simeq\ln(x_1x_2s/\kt^2)$, with $s$ the
squared hadron c.m. energy. Thus there are two
options for increasing $\eta$: performing a {\it ramping run} experiment,
where one fixes $x_1$ and $x_2$ and increases $s$;
increasing $x_1$ and $x_2$ in a fixed-energy collider, like the
Tevatron. The latter is feasible, but introduces a damping in the
cross section, due to the falling parton luminosity \cite{DS}, the former
is desirable because by fixing the $x$'s one fixes as much as possible the
parton densities and looks mainly at the parton dynamics of the function $f$
\cite{MN}, but it requires a variable-energy collider.
A fixed-energy $ep$ collider is nonetheless a variable-energy collider
in the photon-proton frame \cite{muel,bdl}, thus it is possible to increase
$\eta=\ln(x/x_{bj})$ by decreasing $x_{bj}$ while keeping fixed
$x=(\kt e^{\eta_j}/2P)$, with $\eta_j$ the jet rapidity and $P$
the proton energy.

In DIS the gluon-electron cross section with exchange
of a gluon ladder
(\fref{due}) is given in the BFKL approximation of eq.(\ref{solc}) by,
\bea
& & {d\hsig\over dy dQ^2 d\kt^2 d\phi} = {N_c\over\pi^2} \sum_q e_q^2 \a^2
\a_s^2 {1\over (Q^2)^2 \kt^2 y}\label{hot}\\ &\times&
\int {dv_{\perp}^2\over v_{\perp}^2} f(v_{\perp}^2,\kt^2,\tilde{\phi},\eta)
{\cal F}(v_{\perp}^2,Q^2,\hat{\phi},y)\, ,\nn
\eea
with $y$ the electron-energy loss; $Q^2$ and $\qt$ the photon virtuality
and transverse momentum, with $\qt^2 = (1-y)\, Q^2$; $\kt$ and $v_{\perp}$
respectively the transverse momenta of the forward jet and of the gluon
attaching to the quark box; $e_q$ the charge in the quark box;
$\hat{\phi}$ the azimuthal angle between the
photon and the gluon; $\phi$ the azimuthal angle between the outgoing
electron and the jet, with $\phi=\hat{\phi}+\tilde{\phi}+\pi$;
${\cal F}(v_{\perp}^2,Q^2,\hat{\phi},y)$ the impact factor describing the
quark box in the high-energy limit; and with the sum over the final-state
quark flavors.

The cross section for the inclusive forward-jet
production is then given by convoluting the parton cross section (\ref{hot})
with the parton density. Using the explicit form of the impact factor $\cal F$
for final-state light quarks in the HERA lab frame \cite{bdw}, the
production rate is:
\bea
& & {d\sigma\over dy dQ^2 dx d\kt^2 d\phi} = \nn\\ & & f_{eff}(x,\mu)
{N_c\over 8\pi} \sum_q e_q^2 \a^2 \a_s^2 {1\over (Q^2\kt^2)^{3/2} y} \nn\\
&\times& \int_0^{\infty} d\nu\, \cos\left(\nu\ln{Q^2\over\kt^2}\right)\,
{\sinh(\pi\nu)\over\cosh^2(\pi\nu)}\, {1\over\nu(1+\nu^2)} \label{jet}\\
&\times& \left(e^{\omega(\nu,0)\eta}\,\left[\left(3\nu^2+{11\over 4}\right)
(1-y) + \left(\nu^2+{9\over 4}\right){y^2\over 2}\right]\right. \nn\\
&-& \left. e^{\omega(\nu,2)\eta}\, \cos(2\phi)\, \left(\nu^2+{1\over 4}\right)
(1-y)\right)\, ,\nn
\eea
where the integral over $v_{\perp}$ (cf. eq.(\ref{hot})) has been
performed, and where it has been used the effective parton density
\cite{CM}
\beq
f_{eff}(x,\mu) = G(x,\mu) + {4\over 9}\sum_f
\left[Q_f(x,\mu) + \bar Q_f(x,\mu)\right]\, ,\label{effec}
\end{equation}
with the sum over the quark flavors of the incoming parton. Note that in
eq.(\ref{jet}) the dependence on $\phi$ is induced by the impact factor
$\cal F$, and smeared by the gluon ladder; conversely, in the
photon-proton frame the impact factor $\cal F$  does not have any dependence
on $\hat\phi$ in the high-energy limit \cite{bdl}, accordingly in that frame
the production rate (\ref{jet}) has no dependence on $\phi$.

However, when we
integrate out or average on $\phi$, the $\phi$-dependent term drops out.
Then eq.(\ref{jet}) reduces to the production rate
in the photon-proton frame \cite{bdl}. This is because $y$ and $Q^2$ are
Lorentz invariants, and $x$ and $\kt$ are invariant under boosts between the
lab frame and the photon-proton frame in the high-energy limit.

By comparing the relationship between the usual DIS
cross section and the structure functions $F_{1(2)}$,
\beq
{d\s\over dy dQ^2} = {4\pi \a^2\over y Q^4}\, \left[(1-y) F_2(x_{bj},
Q^2) + x_{bj} y^2 F_1(x_{bj}, Q^2)\right]\, ,\label{otto}
\eeq
and eq.(\ref{jet}), with the jet kinematic variables integrated out, we can
compute the contribution of the forward-jet production to the structure
functions $F_{1(2)}$. We perform then a saddle-point evaluation of the
integral over $\nu$, and compute the ratio,
\beq
R^{DIS}(x_{bj},Q^2) = {\s_L(\gamma^* P)\over\s_T(\gamma^* P)} = {1\over
2x_{bj}}\, {F_2(x_{bj},Q^2)\over F_1(x_{bj},Q^2)} - 1\, ,\label{cal}
\eeq
which gives the violations to the Callan-Gross relation, and which for the
forward-jet contribution to the structure functions $F_{1(2)}$ turns out to be
$R^{DIS}(x_{bj},Q^2) = 2/9$.

In performing
a numerical evaluation of the production rate (\ref{jet}) we scale
$\a_s=\a_s(\kt^2)$ from $\a_s(m_Z^2)=0.12$ using the one-loop evolution
with five flavors, fix the factorization scale $\mu$ at $\mu=\kt$, and use
the lowest-order evolved CTEQ parton densities \cite{cteq}.
We define then the rapidity as positive in the proton direction, and take
the acceptance cuts used by the H1 Collaboration \cite{h1}; namely for the
electron $-2.79 \le \eta_e \le -1.59$ and $0.05 \le y
\le 0.55$, from which we obtain $Q^2\ge 4.84\, {\rm GeV}^2$; for the jet
$1.74\le\eta_j\le 2.95$ and $\kt\ge 5\, {\rm GeV}$, from which we obtain
$x\ge 0.017$; in order then to keep the transverse momenta balanced as
required by the multi-Regge kinematics (\ref{mreg}), we select events with
$0.5\le (\kt^2/Q^2)\le 4$. In \fref{tre} we integrate the production
rate (\ref{jet})
over the acceptance cuts, we span $x_{bj}$ over the range $2\cdot 10^{-4} \le
x_{bj}\le 2\cdot 10^{-3}$, and normalize the cross section to the
largest-$x_{bj}$ bin, in order to minimize the normalization errors of the
BFKL approximation. The solid curve is the jet production
rate (\ref{jet}) with $\eta=\ln(x/x_{bj})$, the dashed curve is the production
rate in the Born approximation to the BFKL ladder, which is obtained by
taking the limit $\a_s\eta\ra 0$ in eq.(\ref{jet}). Note that we span a
range of $\eta$ that goes from $\eta\gsim 4.5$ for the smallest-$x_{bj}$
bin to $\eta\gsim 2.2$ for the largest-$x_{bj}$ bin, and we expect the BFKL
approximation to do better in the smallest-$x_{bj}$ bin.

\ffig{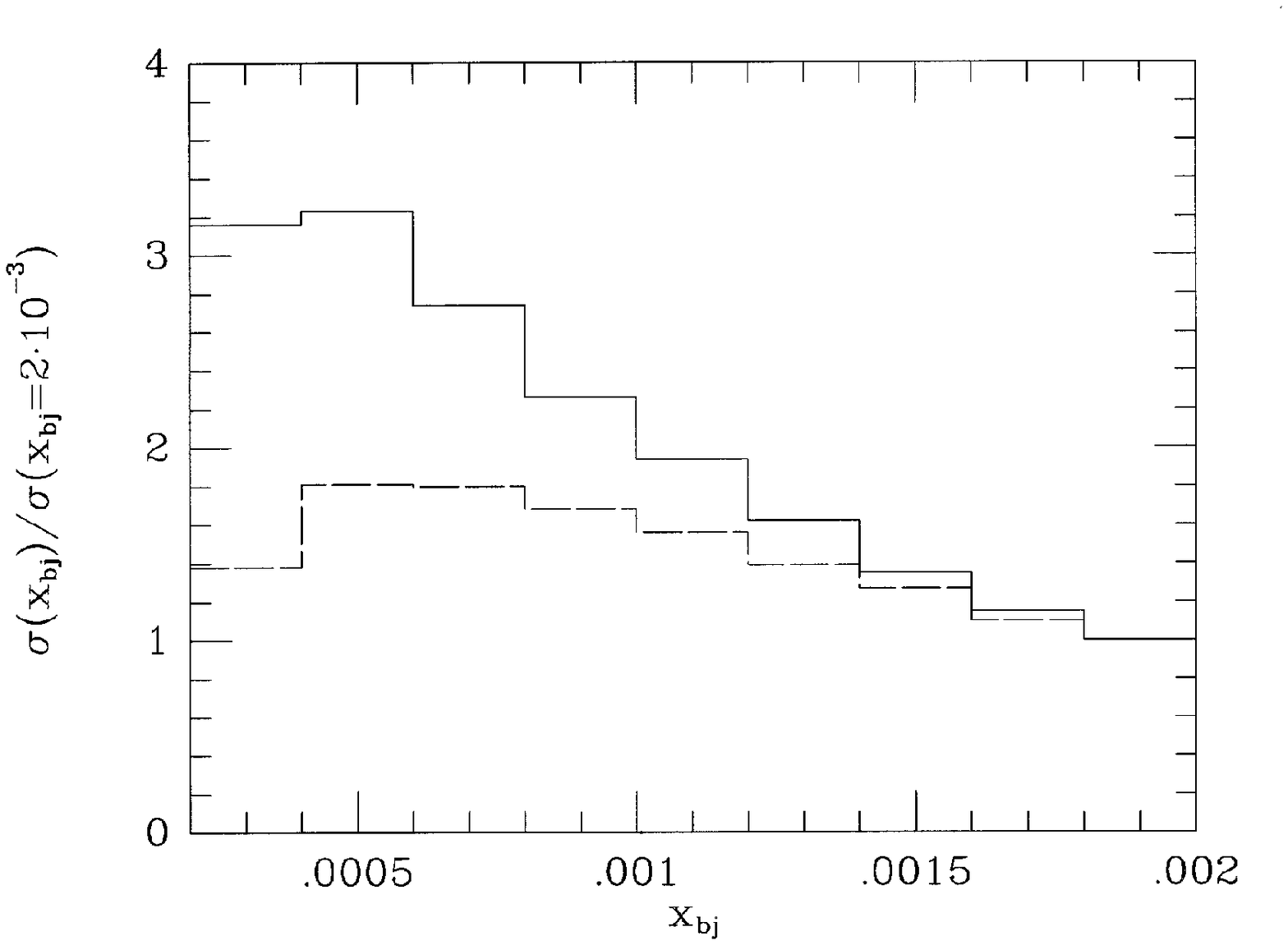}{60mm}{\em Forward-jet production rate, normalized to the
largest-$x_{bj}$ bin, with the BFKL ladder (solid curve), and with the
Born approximation to the BFKL ladder (dashed curve). The acceptance cuts
for the electron and jet kinematic variables are given in the text.}{tre}

A few comments are here
in order: in the Born approximation to the BFKL ladder, the gluon-electron
scattering of \fref{due} reduces to a tree-level diagram with three final-state
partons, since the ladder just reduces to the gluon propagator. The
corresponding forward-jet cross section is infrared finite in the high-energy
limit. In fixed-order perturbation theory the jet cross section based on
the diagram of \fref{due} without the ladder has a collinear divergence,
which is canceled only after adding the 1-loop diagram with two
final-state partons. However in the high-energy limit the diagram with two
final-state partons can not have a gluon exchange
in the crossed channel, thus it does not yield a leading contribution.
Accordingly also the infrared-divergent part of the diagram of \fref{due} is
suppressed.

Next, we consider the azimuthal-angle $\phi$ correlation between the
electron and the forward jet. In two-jet production at large rapidity
intervals, the $\phi$ correlation between the tagging jets has
been predicted to decrease as the rapidity difference, $\Delta\eta$,
between the tagging jets increases \cite{DS}. This phenomenon has been
observed by the D0 Collaboration at the Tevatron Collider \cite{d0}, even
though at this moment it is not clear whether the decorrelation is due to
the evolution in the BFKL ladder, or to the usual DGLAP evolution. In
jet production in DIS
we know that at the parton-model level, i.e. at $x=x_{bj}$, the jet and
the electron are produced back-to-back, and we expect that when $x > x_{bj}$,
but with $\eta=\ln(x/x_{bj})$ still small, the jet production is dominated at
the parton level by the photon-gluon fusion diagram, which has two final-state
partons and is expected to yield the usual correlation at $\phi=\pi$
between the electron and the one of the two partons tagged as the jet.
However as $\eta$ grows the jet production is increasingly
dominated by diagrams with three-final state partons and with gluon
exchange in the crossed channel, and eventually by the higher-order
corrections to them induced by the BFKL ladder (\fref{due}).

Thus we plot in \fref{four} the azimuthal-angle distribution, $N(\phi)$,
as a function of $\phi$, with $0\le\phi\le\pi$, normalized in such a way
that $\int_0^{\infty} d\phi\, N(\phi) = 1$. There four curves are shown: the
\ffig{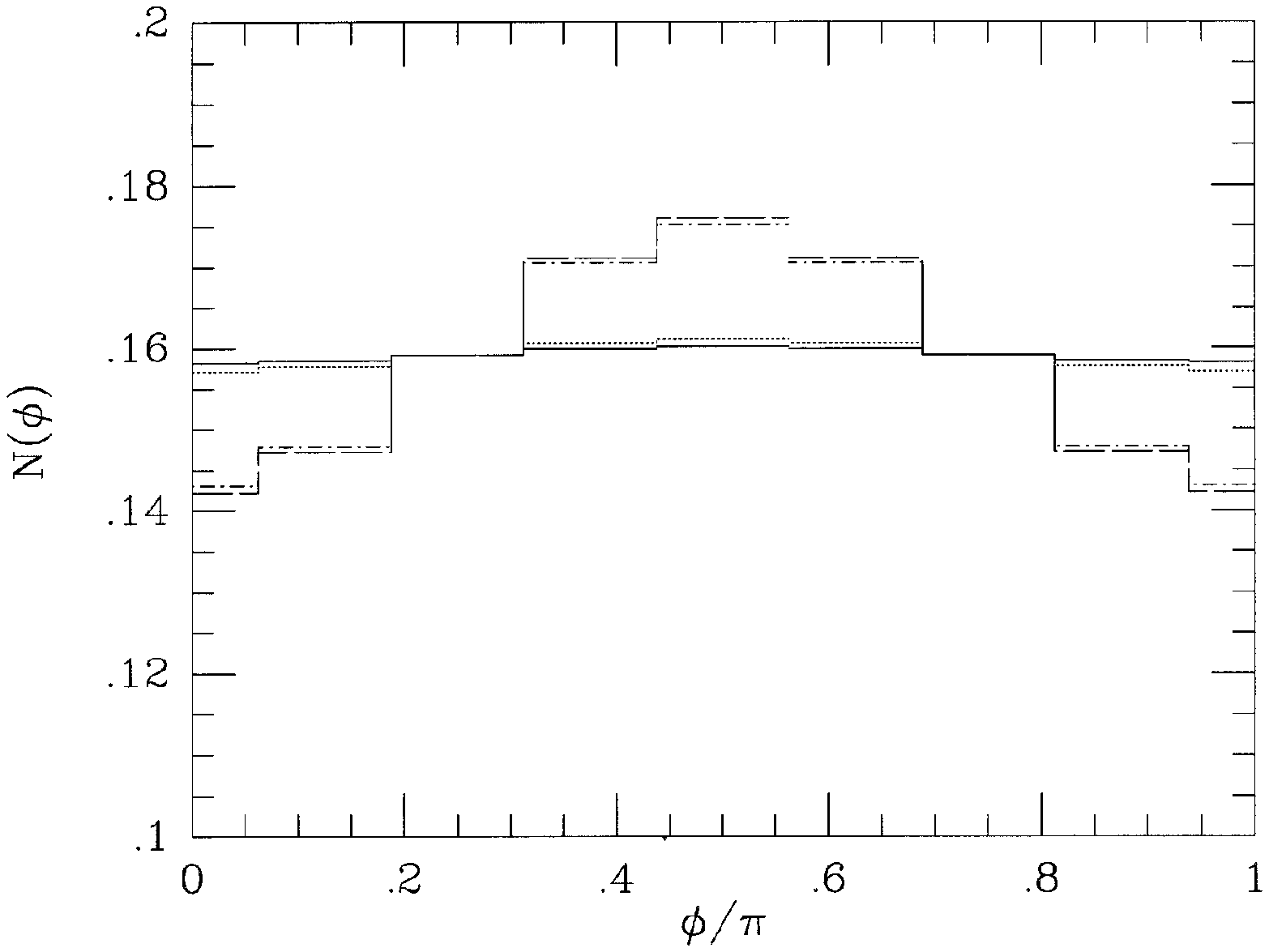}{60mm}{\em Azimuthal-angle distribution, normalized to
the unity. The solid and dotted curves are the distributions for the
forward-jet production with the BFKL ladder, respectively for the bins
$2\cdot 10^{-4} \le x_{bj}\le 10^{-3}$ and $10^{-3} \le x_{bj}\le 2\cdot
10^{-3}$; the dot-dashed and the dashed curves are the distributions with
the Born approximation to the BFKL ladder, respectively for the bins
$2\cdot 10^{-4} \le x_{bj}\le 10^{-3}$ and $10^{-3} \le x_{bj}\le 2\cdot
10^{-3}$. The acceptance cuts for the electron and jet kinematic
variables are the same as in figure~3.}{four}
solid and dotted curves are the distributions computed from
eq.(\ref{jet}) with $\eta=\ln(x/x_{bj})$ for the bins
$2\cdot 10^{-4} \le x_{bj}\le 10^{-3}$ and $10^{-3} \le x_{bj}\le 2\cdot
10^{-3}$ respectively; the dot-dashed and the dashed curves are the
distributions with the Born approximation to the BFKL ladder for
the bins $2\cdot 10^{-4} \le x_{bj}\le 10^{-3}$ and $10^{-3} \le x_{bj}\le
2\cdot 10^{-3}$ respectively.

The distributions $N(\phi)$ are basically
independent of $x_{bj}$; in addition, we note from eq.(\ref{jet}) and the plot
that they are periodic in $\pi$ and peaked at $\phi=\pi/2$.
The correlation at $\phi=\pi/2$ is due to the impact factor $\cal F$,
although we have not been able to track its origin, since in eq.~(\ref{hot})
and (\ref{jet}) we have already integrated over the kinematic variables of
the final-state quarks in the
quark box. This correlation is noticeable for the curves with the Born
approximation to the BFKL ladder, and is basically absent for the ones
with the BFKL ladder.

The Born approximation to the BFKL ladder has a limited range of validity,
since at small $\eta$'s the lower-order photon-gluon fusion diagrams, which
are absent in our analysis, are expected to dominate. Conversely, at
large $\eta$'s the higher-order corrections modeled by the BFKL ladder,
which decorrelate the lepton and the jet, should become important.
Therefore we may envisage a three-fold scenario for the distribution $N(\phi)$:
\begin{itemize}
\item[(i)] at small $\eta$'s, the lepton and the jet are back-to-back, i.e.
there is a correlation at $\phi=\pi$, yielded by the jet production rate
at $O(\a^2\a_s)$, i.e. by the photon-gluon fusion diagrams at the parton level;
\item[(ii)] at intermediate $\eta$'s, we have the correlation at $\phi=\pi/2$
(figure 4), yielded by the jet production rate at $O(\a^2\a_s^2)$, i.e. by
the Born approximation to the BFKL ladder in figure 2;
\item[(iii)] at large $\eta$'s, the BFKL ladder decorrelates the lepton and the
jet (figure 4).
\end{itemize}
For the acceptance cuts of figures~3 and 4 the range of $\eta$ is $2.2\le
\eta\le 4.5$. We cannot, though, unambiguosly compare this range to
the scenario above because we have no control on the theoretical error in
our analysis. Thus it would be desirable to compare our analysis
to a complete calculation at $O(\a^2\a_s^2)$, on the theoretical side,
and of course to data.


\begin{center}
{\large\bf Aknowledgements}
\end{center}
Thanks to Albert de Roeck for providing the acceptance
cuts used in the analysis of the H1 Collaboration; I would also like
to acknowledge the hospitality of the CERN Theory
Group where part of this work was completed.
\vspace{2cm}


\Bibliography{100}

\bibitem{lip} L.N.~Lipatov, Yad.~Fiz. {\bf 23}, 642 (1976)
[Sov.~J.~Nucl.~Phys. {\bf 23}, 338 (1976)];
E.A.~Kuraev,~L.N.~Lipatov~and~V.S.~Fadin, Zh.~Eksp.~Teor.~Fiz. {\bf 71}, 840
(1976) [Sov.~Phys.~JETP {\bf 44}, 443 (1976)]; Zh.~Eksp.~Teor.~Fiz. {\bf 72},
377 (1977) [Sov.~Phys.~JETP {\bf 45}, 199 (1977)];
Ya.Ya.~Balitsky and L.N.~Lipatov, Yad.~Fiz. {\bf 28} 1597 (1978)
[Sov.~J.~Nucl.~Phys. {\bf 28}, 822 (1978)].

\bibitem{cat} S.~Catani, in these proceedings.

\bibitem{MN} A.H.~Mueller and H.~Navelet, Nucl.~Phys. {\bf B282}, 727 (1987).

\bibitem{DS} V.~Del~Duca and C.R.~Schmidt, Phys.~Rev.~D {\bf 49}, 4510 (1994);
Phys.~Rev.~D {\bf 51}, 2150 (1995); Nucl.Phys. {\bf B} (Proc.Suppl.) {\bf 39C},
137 (1995); W.J.~Stirling, Nucl.~Phys. {\bf B423}, 56 (1994).

\bibitem{d0} T.~Heuring for the D0 Collab., talk at the ${\rm XXX}^{th}$
Rencontres de Moriond on ``QCD and High Energy Hadronic Interactions",
France, March 1995.

\bibitem{muel} A.H.~Mueller, Nucl.Phys. {\bf B} (Proc.Suppl.) {\bf 18C}, 125
(1991).

\bibitem{bdl} J.~Bartels, A.~De~Roeck and M.~Loewe, Z.~Phys. {\bf C54}, 635
(1992); W.-K.~Tang, Phys.~Lett. {\bf 278B}, 363 (1991).

\bibitem{bdw} J.~Bartels, V.~Del~Duca and M.~W\"usthoff, in preparation.

\bibitem{CM} B.L.~Combridge and C.J.~Maxwell, Nucl.~Phys. {\bf B239}, 429
(1984).

\bibitem{cteq} J.~Botts et al., Phys.~Rev.~D {\bf 51}, 4763 (1995).

\bibitem{h1} T.~Ahmed et al., H1 Collab., preprint DESY 95-108.

\end{thebibliography}
\end{document}